\documentclass[aps,superscriptaddress,eqsecnum,nofootinbib,preprintnumbers]{revtex4}
\usepackage{graphicx,epsfig}
\usepackage{amsmath}
\usepackage {amssymb}
\usepackage{subfigure}

\usepackage{relsize}
\newcommand{\be}{\begin{equation}}
\newcommand{\ee}{\end{equation}}
\newcommand{\bea}{\begin{eqnarray}}
\newcommand{\eea}{\end{eqnarray}}
\newcommand{\nn}{\nonumber}

\begin{document}

\title{Modified Brans-Dicke cosmology with matter-scalar field interaction}

\author{Georgios Kofinas}
\email{gkofinas@aegean.gr} \affiliation{Research Group of Geometry,
Dynamical Systems and Cosmology,
Department of Information and Communication Systems Engineering\\
University of the Aegean, Karlovassi 83200, Samos, Greece}

\author{Eleftherios Papantonopoulos}
\email{lpapa@central.ntua.gr} \affiliation{Physics Division,
National Technical University of Athens, 15780 Zografou Campus,
Athens, Greece}

\author{Emmanuel N. Saridakis}
\email{Emmanuel\_Saridakis@baylor.edu}
\affiliation{CASPER, Physics Department, Baylor University, Waco, TX 76798-7310, USA}
\affiliation{Instituto de F\'{\i}sica, Pontificia
Universidad de Cat\'olica de Valpara\'{\i}so, Casilla 4950,
Valpara\'{\i}so, Chile}

\begin{abstract}

We discuss the cosmological implications of an extended
Brans-Dicke theory presented recently, in which there is an energy exchange between
the scalar field and ordinary matter, determined by the theory. A new mass scale is
generated in the theory which modifies the  Friedmann equations
with field-dependent corrected kinetic terms. In a radiation
universe the general solutions are found and there are branches
with complete removal of the initial singularity,  while at the
same time a transient accelerating period can occur within
deceleration. Entropy production is also possible in the early universe.
In the dust era, late-times acceleration has been
found numerically in agreement with the correct behaviour of the
density parameters and the dark energy equation of state, while
the gravitational constant has only a slight variation over a
large redshift interval in agreement with observational bounds.

\end{abstract}

\maketitle

\section{Introduction} \label{Introduction}

Scalar fields can play an important role in the description of the
early universe as well as the late-times cosmic evolution.
Scalar-tensor gravitational theories are widely studied as an
alternative to General Relativity. Brans-Dicke (BD) gravity
\cite{Brans:1961sx} is one of the simplest such theories that can
be constructed and it is considered a viable alternative to
General Relativity, one which respects Mach's principle and weak
equivalence principle. Mach's principle states that the property
of inertia of material bodies arises from their interactions with
the matter distributed in the universe. This theory leads to
variations in the Newtonian gravitational constant $G$ and
introduces a new dimensionless coupling constant $\omega$. Large
values of $\omega$ mean a significant contribution from the tensor
part, while small values of $\omega$ mean an increasing role for
the scalar field contribution. General Relativity is recovered in
the limit $\omega\rightarrow \infty$. The effective gravitational
constant in Brans-Dicke theory is the inverse of the scalar field, namely
$G\sim 1/\phi$, however from spherically symmetric solutions it
is $G=\frac{4+2\omega}{3+2\omega}\frac{1}{\phi}$.

In modern context, BD theory appears naturally in supergravity models, from
string theory at low-energies in the so-called string (Jordan)
frame or from the dimensional reduction of Kaluza-Klein theories
\cite{Freund:1982pg}, and this has led to considerable interest in
BD gravity. This theory yields the correct Newtonian weak-field
limit, but solar system measurements of post-Newtonian corrections
require that $\omega$ is larger than a few thousands
\cite{Bertotti:2003rm}. This is due to the fact that in order to
avoid the propagation of the fifth force, the coupling between
matter and the massless field $\phi$ should be suppressed. On the
other hand, from cosmological observations  $\omega$ gets
substantially lower values in a model dependent way
\cite{Acquaviva:2004ti}. The synthesis of light elements during the early Universe
\cite{green} provides extra observational constraints upon
scalar-tensor theories. There is also the possibility that the gravitational
coupling depends on the scale \cite{Mannheim:1999bu}, having
different value at local and at cosmological scale, and thus it is
possible $\omega$ to be smaller at cosmological scales giving
deviations from General Relativity, while agreement with local
tests is preserved.

Many cosmological observations from type Ia supernovae, cosmic microwave background
radiation and large scale structure reveal that our universe is undergoing an accelerating
expansion. The mysterious component with large enough negative pressure which dominates
the dynamics of the universe and drives the cosmic acceleration is called dark energy. The
preferred candidate of dark energy is the Einstein's cosmological constant which fits the
observations well, but is plagued by the fine-tuning and the cosmic coincidence problems.
This recent accelerating period has to be replaced in the past by a decelerating era in
order to accommodate for nucleosynthesis in the radiation era and for the formation of
galaxies in the matter era.

Most of the models in cosmology consider that the evolutions of
dark matter and dark energy occur separately. In
\cite{Bertolami:2007zm} it is argued that observational evidences
support an interaction between dark energy and dark matter and
violation of equivalence principle between baryons and dark
matter. Recently there is a raising activity in interacting models
of dark matter and dark energy \cite{Wang:2006qw}, and this flow
of energy between the dark matter and the dark energy component
can be useful to solve the coincidence problem
\cite{Amendola:1999qq}. Possible mechanisms could alleviate the puzzles arising
by the violation of the equivalence principle. In Chameleon mechanism
\cite{Khoury:2003aq} the effective mass of the scalar field can
become density dependent, so a large effective mass may be
acquired in solar scale hiding local experiments, while at
cosmological scales $\phi$ can be effectively light providing
cosmological modifications. An alternative for the resolution of
this problem and the recovery of General Relativity in the regions
of high energy is the introduction of non-linear self interactions
for $\phi$ through self screening (Vainshtein) mechanisms
\cite{Vainshtein:1972sx}. Finally, since the validity of the
universality of free fall at cosmological scales has not been
tested directly, there is the option that the baryonic matter is
separately conserved, so as to obey the weak equivalence principle,
while dark matter interacts with dark energy.

A series of papers  \cite{nariai}-\cite{Bertobarrow} have found
cosmological solutions of Brans-Dicke gravity for radiation, dust
or other equations of state or in the presence of a cosmological
constant and for all kinds of spatial curvature. Although in
cosmology the evolution of the universe generically has a
singularity in the past, usually of big bang type as in standard
cosmology, in \cite{Park:1997dw} it was argued that a gas of
solitonic $p-$branes treated as a perfect fluid type matter can
resolve the initial singularity in the Brans-Dicke theory. In
\cite{Tretyakova:2011ch} bouncing solutions were found for
negative $\omega$ in dust-filled universe, while dynamical systems
analysis of FRW cosmologies was given in \cite{Kolitch:1994qa}.
Brans-Dicke theory is useful in solving some problems of the
inflationary scenario with the possibility of extended inflation
\cite{Mathiazhagan:1984vi}-\cite{Linde:1989tz}. The solution to
the ``graceful exit'' problem of inflation in terms of an extended
inflation scenario \cite{La:1989za} was first obtained in
Brans-Dicke theory without fine tuning, although the value of
$\omega$ is small in order not to create large anisotropies in the
microwave background (see also \cite{Berkin:1991nm} where the
introduction of a potential for the scalar field or a scalar field
dependent $\omega$ were proposed to solve such problems).

An attractive feature of BD theory is that the scalar field is a
fundamental element of the theory which controls the evolution of
the gravitational constant and at the same time may possibly form
the dark energy. However, it is difficult to succeed acceleration
in the standard version of Brans-Dicke theory. There have been
found accelerating solutions with $-2\leq\omega\leq -3/2$ in the
matter-dominated universe for a spatially flat universe without
cosmological constant or quintessence \cite{Banerjee:2000mj}. Such
small values of $|\omega|$ not only are not in agreement with
solar system constraints, they also violate the energy conditions
on the scalar field and they do not provide a transition to a
decelerating era (see also \cite{Batista:2000iy}). Spatially
closed models with higher values of $\omega$ were considered in
\cite{Chakraborty:2008xt}.

Since in the limit $\omega\rightarrow \infty$ the field $\phi$
becomes fixed and we recover Einstein gravity, this has led to the
construction of more general scalar-tensor gravity theories with a
self-interacting potential $V(\phi)$, a field or time-dependent
$\omega$, or non-minimal couplings \cite{Bergmann:1968ve}. The
late-times acceleration of the universe in such models has been
studied in  \cite{Bartolo:1999sq}-\cite{deRitis:1999zn}, and
either fail or succeed to obtain the standard decelerating phase
of the universe followed by the recent accelerating period.
Although the majority of the models refer to the Jordan frame, in
\cite{Sen:2001ki} a suitably selected self-interacting potential
in Einstein conformal frame was selected and found a class of
solutions with accelerated expansion, large positive $\omega$ and
constant ratio of energy densities of matter and scalar field.
There are also other modifications of Brans-Dicke cosmology, such
as the introduction of a dissipative cold dark matter fluid
\cite{Sen:2000hg}, a transfer of energy between the dark matter
and the Brans-Dicke scalar field \cite{Das:2005yg} (or in the
presence of Chaplygin gas \cite{Sahoo:2002rx}), the introduction
of a chameleon field \cite{Banerjee:2008rs}, holographic dark
energy models in the framework of Brans-Dicke cosmologies
\cite{Chen:2003ed}, or other scenarios \cite{Chakraborty:2003ye}.

Recently, relaxing  the standard exact conservation of the matter
energy-momentum, but still preserving the simple massless wave equation of
motion for the scalar field sourced by the trace of the matter energy-momentum
tensor, three general completions of Brans-Dicke gravity were found which are
uniquely determined from the consistency of the Bianchi identities \cite{Kofinas:2015nwa}.
Here, we will focus on the first of these theories, where a new dimensionfull parameter
$\nu$ appears (which is an integration constant) and for $\nu=0$ it reduces to the standard
Brans-Dicke theory. The energy transfer between ordinary matter and the scalar
field defines the non-conservation
equation of motion of the matter. The derivation of the theory
emerged initially at the level of the field equations. A discussion on the action of the
theory was given in \cite{Kofinas:2015sjz}, where the vacuum action was derived, as well
as the full action for special only matter Lagrangians. The
matter Lagrangian due to the interaction turns out to be non-minimally coupled
even in the Jordan frame, while the issue of the general action of the full theory still remains open.

In the present work we study the cosmological evolution of this
theory at early and late times. In the radiation regime we find
general exact solutions depending on the parameters of the theory.
In the dust period we investigate numerically the cosmic
possibilities, where the main mechanism is the energy exchange
between dark matter and dark energy predicted by the theory. The
scalar field plays at the same time the role of the varying
gravitational constant and the role of dark energy. In both cosmic
eras we find interesting behaviours of the cosmic evolution which
could imitate the history of the universe. The really tempting
feature in our approach is the fact that we are strictly focused
on the extended Brans-Dicke theory which possesses only
well-defined kinetic terms, and we have not added any ad-hoc
structures such as self-interacting potentials or varying
functions, contrary to the standard Brans-Dicke cosmology where
some extra ingredient is necessary to make the model viable.

The content of the paper is organized as follows: In section II we write down and
elaborate the new cosmological equations and integrate the new conservation equation.
In section III we integrate the system in the radiation epoch, what leads to the general
solutions of the scale factor as a function of the scalar field and discuss their
implications at early times. In section IV we numerically integrate the system in the dust
regime, we confront against the basic observational data and find the possibility of a
recent accelerating era following the decelerating period. Finally, in section V we
summarize our results and conclude.

\section{Cosmological equations}
The standard Brans-Dicke theory is described by the equations (in units where the velocity
of light is set to 1)
\label{cosmo}
\begin{eqnarray}
G^{\mu}_{\,\,\,\nu}\!\!&=&\!\!\frac{8\pi}{\phi}(T^{\mu}_{\,\,\,\nu}+
\mathcal{T}^{\mu}_{\,\,\,\,\nu})
\label{kns}\\
T^{\mu}_{\,\,\,\nu}\!\!&=&\!\!\frac{2-3\lambda}{16\pi\lambda\phi}\Big(
\phi^{;\mu}\phi_{;\nu}\!-\!\frac{1}{2}\delta^{\mu}_{\,\,\,\nu}
\phi^{;\rho}\phi_{;\rho} \Big)
\!+\!\frac{1}{8\pi}\big(\phi^{;\mu}_{\,\,\,\,;\nu}\!-\!\delta^{\mu}_{\,\,\,\nu}
\Box\phi\big)
\label{qqd}\\
\Box\phi\!\!&=&\!\!4\pi\lambda\mathcal{T}\label{lwn}\\
\,\,\mathcal{T}^{\mu}_{\,\,\,\,\nu;\mu}\!\!&=&\!\!0~. \label{jrk}
\end{eqnarray}
These equations are resulted from a simple action in the Jordan frame of the form
\begin{equation}
S_{BD}=\frac{1}{16\pi}\int \!d^{4}x \,\sqrt{-g} \,\Big(\phi
R-\frac{\omega}{\phi}g^{\mu\nu}\phi_{,\mu}\phi_{,\nu}\Big)+\int \!d^{4}x
\,\sqrt{-g} \,L_{m}~, \label{jsw}
\end{equation}
where $L_{m}(g_{\kappa\lambda},\Psi)$ is the matter Lagrangian
depending on scalar fields $\Psi$. The parameter $\lambda\neq 0$
is dimensionless and is related to the standard Brans-Dicke
parameter $\omega$ by $\omega=(2-3\lambda)/2\lambda$. The tensor
$\mathcal{T}^{\mu}_{\,\,\,\,\nu}$ is the matter energy-momentum
tensor and $\mathcal{T}=\mathcal{T}^{\mu}_{\,\,\,\,\mu}$ is its
trace.

The standard Brans-Dicke theory described by equations
(\ref{kns})-(\ref{jrk}) was generalized in \cite{Kofinas:2015nwa}
by relaxing the strict conservation law (\ref{jrk}), but still
respecting the simple form of the scalar field equation (\ref{lwn}).
As usual, the energy-momentum tensor of the
scalar field $T^{\mu}_{\,\,\,\,\nu}$ was constructed from terms
each of which involves two derivatives of one or two scalar fields
$\phi$, and $\phi$ itself. Three unique theories were unambiguously determined
from consistency at the level of the equations of motion, and here we study
the first of these theories which has the following form
\begin{eqnarray}
&&\!\!\!\!\!\!\!G^{\mu}_{\,\,\,\nu}\!=\!\frac{8\pi}{\phi}
(T^{\mu}_{\,\,\,\nu}+\mathcal{T}^{\mu}_{\,\,\,\,\nu})
\label{elp}\\
&&\!\!\!\!\!\!\!T^{\mu}_{\,\,\,\nu}\!=\!\frac{\phi}{2\lambda(\nu\!+\!8\pi\phi^{2})^{2}}
\Big{\{}
2\big[(1\!+\!\lambda)\nu\!+\!4\pi(2\!-\!3\lambda)\phi^{2}\big]\phi^{;\mu}\phi_{;\nu}
\!-\!\big[(1\!+\!2\lambda)\nu\!+\!4\pi(2\!-\!3\lambda)\phi^{2}\big]\delta^{\mu}_{\,\,\,\nu
} \phi^{;\rho}\phi_{;\rho} \Big{\}}
\!+\!\frac{\phi^{2}}{\nu\!+\!8\pi\phi^{2}}
\big(\phi^{;\mu}_{\,\,\,\,;\nu}\!-\!\delta^{\mu}_{\,\,\,\nu}\Box\phi\big)
\nn\\
\label{utd}\\
&&\!\!\!\!\!\!\!\Box\phi\!=\!4\pi\lambda\mathcal{T}\label{lrs}\\
&&\!\!\!\!\!\!\!\mathcal{T}^{\mu}_{\,\,\,\,\nu;\mu}\!=\!\frac{\nu}{\phi(\nu\!+\!8\pi\phi^{
2})} \mathcal{T}^{\mu}_{\,\,\,\,\nu}\phi_{;\mu}~. \label{idj}
\end{eqnarray}
The parameter $\nu$ is arbitrary and has dimensions mass to the fourth (it arises as an
integration constant from the integration procedure). Its sign and numerical value should be
determined experimentally.
For $\nu=0$ the above system of equations reduces to the standard Brans-Dicke
theory (\ref{kns})-(\ref{jrk}). The role of $\nu$ is manifest in
equation (\ref{idj}) and measures the deviation from the exact
conservation of matter. The right-hand side of equation
(\ref{elp}) is consistent with the Bianchi identities, i.e. it is
covariantly conserved on-shell, and therefore, the system of
equations (\ref{elp})-(\ref{idj}) is well-defined.

To investigate the cosmological implications of the above theory, we consider the
following spatially homogeneous and isotropic ansatz
\begin{equation}
ds^{2}=-dt^{2}+a(t)^{2}\Big[\frac{dr^{2}}{1\!-\!kr^{2}}+r^{2}\big(d\theta^{2}\!+\!\sin^{2}
\!{\theta} \, d\varphi^{2}\big)\Big]~, \label{metric}
\end{equation}
characterized by the spatial curvature $k=-1,0,1$. The temporal
gauge choice has been assumed with the lapse function being unity,
so $t$ is the cosmic time. The scalar field $\phi$ respecting the
symmetries of the metric (\ref{metric}) will be a function of
time, so it is $\phi(t)$. The matter energy-momentum tensor is the
one of a perfect fluid
$\mathcal{T}^{\mu}_{\,\,\,\,\nu}=\text{diag}(-\rho,p,p,p)$ with
$\rho(t)$ being its energy density and $p(t)$ its pressure.
Equations (\ref{elp}), (\ref{lrs}) and  (\ref{idj}) become
respectively
\begin{eqnarray}
&&H^{2}+\frac{k}{a^{2}}=\frac{8\pi}{3\phi}\rho-\frac{8\pi\phi}{\nu\!+\!8\pi\phi^{2}}H\dot{
\phi}
+\frac{4\pi}{3\lambda}\,\frac{\nu\!+\!4\pi(2\!-\!3\lambda)\phi^{2}}
{(\nu\!+\!8\pi\phi^{2})^{2}}\dot{\phi}^{2}
\label{eji}\\
&&2\dot{H}+3H^{2}+\frac{k}{a^{2}}=-\frac{8\pi}{\phi}\Big[
p+\frac{\phi}{2\lambda}\,\frac{(1\!+\!2\lambda)\nu\!+\!4\pi(2\!-\!3\lambda)\phi^{2}}
{(\nu\!+\!8\pi\phi^{2})^{2}}
\dot{\phi}^{2}+\frac{\phi^{2}}{\nu\!+\!8\pi\phi^{2}}(2H\dot{\phi}+\ddot{\phi})\Big]
\label{ket}\\
&&\ddot{\phi}+3H\dot{\phi}+4\pi\lambda(3p\!-\!\rho)=0
\label{iet}\\
&&\dot{\rho}+3H(\rho\!+\!p)=\frac{\nu}{\phi(\nu\!+\!8\pi\phi^{2})}\rho\,\dot{\phi}~.
\label{euf}
\end{eqnarray}
The consistency of the above system can be confirmed by checking
that the satisfaction of the Bianchi identities is verified. This
results to the fact that one of the dynamical equations is
redundant and arises from the other equations. We also give for
comparison the Brans-Dicke equations which arise from equations
(\ref{eji})-(\ref{euf}) setting $\nu=0$
\begin{eqnarray}
&&H^{2}+\frac{k}{a^{2}}=\frac{8\pi}{3\phi}\rho-H\frac{\dot{\phi}}{\phi}
+\frac{2\!-\!3\lambda}{12\lambda}\frac{\dot{\phi}^{2}}{\phi^{2}}
\label{eew}\\
&&2\dot{H}+3H^{2}+\frac{k}{a^{2}}=-\frac{1}{\phi}\Big(8\pi
p+\frac{2\!-\!3\lambda}{4\lambda\phi}
\dot{\phi}^{2}+2H\dot{\phi}+\ddot{\phi}\Big)
\label{wrg}\\
&&\ddot{\phi}+3H\dot{\phi}+4\pi\lambda(3p\!-\!\rho)=0
\label{erj}\\
&&\dot{\rho}+3H(\rho\!+\!p)=0~.\label{eue}
\end{eqnarray}

Equation (\ref{euf}) for $p=w\rho$ can be integrated to give
\begin{equation}
\rho=\frac{\rho_{\ast}}{a^{3(1+w)}}\frac{|\phi|}{\sqrt{|\nu\!+\!8\pi\phi^{2}}|}~,
\label{keg}
\end{equation}
where $\rho_{\ast}>0$ is an integration constant. Observe that the
energy density evolves in a different way than what would be
expected for a self-conserved matter sector. This happens because
there is a direct coupling of the matter energy density to the
scalar field and this behaviour has important consequences for the
cosmological evolution, as we will discuss in the next sections.
Since $\phi$ is related to the gravitational constant through the
relation
\begin{eqnarray}
G=\frac{1}{\phi}~, \label{Geff}
\end{eqnarray}
normally, it should be $\phi>0$. However, there is a
possibility that at the early stages of the universe evolution it
is $\phi<0$, providing an antigravity effect and possibly
contributing to the inflationary phase. Therefore, we will include
in the following equations both signs of $\phi$ and restrict our
discussion on the properties of the radiation solutions for
$\phi>0$ (of course, $\phi$ is also positive in the dust universe).
Note that a flip from a negative to a positive $\phi$ seems unnatural since
the dynamical evolution of the scalar field would lead to
the undesirable situation of going through an infinitely strong gravitational effect.

The relation (\ref{Geff}) arises from the cosmological
considerations  of the theory. However, the full correspondence
with $\phi$ would arise from the spherical solutions of the theory
and the comparison with the solar system data. However, since in
the present work we are interested in investigating the
cosmological implications of the theory at hand, the
identification (\ref{Geff}) is adequate. Hence, the only
requirement that we should be careful of, is the variance of $G$
to be in agreement with the observational limits, namely
$-0.5\times
10^{-12}\,\text{yr}^{-1}\lesssim\frac{\dot{G}}{G}\big|_{0}\lesssim
1.7\times 10^{-12}\,\text{yr}^{-1}$ \cite{Zhu:2015mdo}, which in
terms of the present value of the Hubble parameter $H_0\approx
(0.71\pm 0.008)\times 10^{- 10} \,\text{yr}^{-1}$
\cite{Ade:2013sjv}, becomes $-0.7\times
10^{-2}\lesssim\frac{\dot{G}}{GH}\big|_{0}\lesssim2.4\times
10^{-2}$. This constraint, due to (\ref{Geff}), becomes $-2.4\times
10^{-2}\lesssim\frac{\dot{\phi}}{\phi H}\big|_{0}\lesssim0.7\times
10^{-2}$. Finally, note that the variation of $G$ is also
constrained during the early cosmology by the light elements
production in Big Bang Nucleosyntesis (BBN) epoch, with the
corresponding bound reading as $\frac{|\delta {G}|}{G_0
H_0}\lesssim 0.2$ where $\delta G=G_{BBN}-G_0$
\cite{Bambi:2005fi}, which is not a rather strong constraint.
No other use of the relation (\ref{Geff}) will be made in this work, except from the
satisfaction of the above bounds on the time variability of the scalar field.

Before we finish this section we can extract the acceleration
$\frac{\ddot{a}}{a}$ which will be useful in the following,
combining equations (\ref{eji})-(\ref{iet}) to be
\begin{equation}
\frac{\ddot{a}}{a}=-\frac{4\pi\rho}{\phi}\Big[w\!+\!\frac{1}{3}\!+\!\lambda(1\!-\!3w)
\frac{4\pi\phi^{2}}{\nu\!+\!8\pi\phi^{2}}\Big]+\frac{8\pi\phi}{\nu\!+\!8\pi\phi^{2}}H\dot{
\phi}
-\frac{4\pi}{3\lambda}\frac{(2\!+\!3\lambda)\nu\!+\!8\pi(2\!-\!3\lambda)\phi^{2}}
{(\nu\!+\!8\pi\phi^{2})^{2}} \dot{\phi}^{2}~, \label{jwt}
\end{equation}
or also
\begin{equation}
\frac{\ddot{a}}{a}=-H^{2}-\frac{k}{a^{2}}+\frac{4\pi(1\!-\!3w)\rho}{3\phi}
\Big(1\!-\!\frac{12\pi\lambda\phi^{2}}{\nu\!+\!8\pi\phi^{2}}\Big)-\frac{4\pi}{3\lambda}\,
\frac{(1\!+\!3\lambda)\nu\!+\!4\pi(2\!-\!3\lambda)\phi^{2}}{(\nu\!+\!8\pi\phi^{2})^{2}}
\dot{\phi}^{ 2}~. \label{egs}
\end{equation}

Finally, note that the case with $\nu\!+\!8\pi\phi^{2}>0$ includes two subcases, the first
one
with $\nu>0$ and the second with $\nu<0$, $|\phi|>\sqrt{\frac{|\nu|}{8\pi}}$. The case
with
$\nu\!+\!8\pi\phi^{2}<0$ corresponds to $\nu<0$, $|\phi|<\sqrt{\frac{|\nu|}{8\pi}}$.

\section{Early-times cosmological evolution}

We will study the cosmological evolution when the universe is in
the radiation era defined by the relativistic equation of state
$w=1/3$. Then, from the scalar field equation (\ref{iet}) we see that the scalar field
evolution is decoupled from the radiation and this equation is integrated to
\begin{equation}
\dot{\phi}a^{3}=c~, \label{drs}
\end{equation}
where $c$ is an integration constant. From this equation we get
\begin{equation}
H=\frac{c}{a^{4}}\frac{da}{d\phi}~. \label{erg}
\end{equation}
It is seen from equation (\ref{drs}) that $\phi$ is directly
connected with the scale factor, so $\phi(t)$ is either a
monotonically increasing or decreasing function in the radiation
era. Substituting the expressions (\ref{keg}), (\ref{drs}) and
(\ref{erg}) into the Friedmann equation (\ref{eji}) we get an
equation for  $a(\phi)$
\begin{equation}
\Big(\frac{da}{d\phi}\!+\!\frac{4\pi\phi
a}{\nu\!+\!8\pi\phi^{2}}\Big)^{\!2}
-\frac{4\pi}{3\lambda}\,\frac{a^{2}}{\nu\!+\!8\pi\phi^{2}}\Big(1\!+\!
\frac{2\epsilon\lambda\rho_{\ast}}
{c^{2}}a^{2}\sqrt{|\nu\!+\!8\pi\phi^{2}|}\Big)+\frac{ka^{6}}{c^{2}}=0~,
\label{weg}
\end{equation}
where $\epsilon=\text{sgn}[\phi(\nu\!+\!8\pi\phi^{2})]$.
Defining
\begin{equation}
\chi=a^{2}\sqrt{|\nu\!+\!8\pi\phi^{2}|}~, \label{erq}
\end{equation}
we find from equation (\ref{weg})
\begin{equation}
\Big(\frac{d\chi}{d\phi}\Big)^{\!2}-\frac{16\pi}{3\lambda}\,\frac{\chi^{2}}{
\nu\!+\!8\pi\phi^{2}}
\Big(1\!+\!\frac{2\epsilon\lambda\rho_{\ast}}{c^{2}}\chi\Big)+\frac{4k}{c^{2}}
\frac{\chi^{4}}{|\nu\!+\!8\pi\phi^{2}|}=0~. \label{wet}
\end{equation}
Equation (\ref{wet}) is a separable equation and can be solved for any $k$. However, for
simplicity
we will restrict our interest here to $k=0$. Then, equation (\ref{wet}) becomes
\begin{equation}
\frac{d\chi}{d\phi}=\pm 4\sqrt{\frac{\pi}{3}}\frac{\chi}{\sqrt{|\nu\!+\!8\pi\phi^{2}|}}
\sqrt{\frac{\epsilon}{\lambda}\!+\!\frac{2\rho_{\ast}}{c^{2}}\chi}\,,
\label{rta}
\end{equation}
where
$\frac{\epsilon}{\lambda}\!+\!\frac{2\rho_{\ast}}{c^{2}}\chi$ has
to be non-negative. There are four cases for the integration of
(\ref{rta}) concerning the sign of $\nu\!+\!8\pi\phi^{2} $ and the
sign of $\epsilon\lambda$. Before presenting the solutions of
equation (\ref{rta}), we give the expressions for the time
integral, the acceleration and the curvature scalar.

The time dependence of $\phi$ is found from equation (\ref{drs})
as
\begin{equation}
t=\frac{1}{c}\int\! a(\phi)^{3}d\phi\,,
\label{erv}
\end{equation}
where a translational integration constant for time has been
absorbed. The acceleration $\frac{\ddot{a}}{a}$ is found from
(\ref{jwt}) to be
\begin{equation}
\frac{\ddot{a}}{a}=-\frac{8\pi\rho}{3\phi}+\frac{8\pi\phi}{\nu\!+\!8\pi\phi^{2}}H\dot{\phi
}
-\frac{4\pi}{3\lambda}\frac{(2\!+\!3\lambda)\nu\!+\!8\pi(2\!-\!3\lambda)\phi^{2}}
{(\nu\!+\!8\pi\phi^{2})^{2}} \dot{\phi}^{2}\,. \label{jwp}
\end{equation}
In the above equation, the quantity $\frac{\rho}{\phi}$ is
obtained from equation (\ref{keg}); the Hubble parameter $H$ is
obtained from equation (\ref{erg}), where $\frac{da}{d\phi}$ is
found from equation (\ref{weg}); finally, the quantity
$\dot{\phi}$ is obtained from equation (\ref{drs}). Thus,
$\frac{\ddot{a}}{a}$ becomes a function of $a,\phi$ and after
integration of (\ref{rta}), it becomes a function of only $\phi$.
From equation (\ref{egs}) we have also the expression
\begin{equation}
\frac{\ddot{a}}{a}=-H^{2}-\frac{4\pi}{3\lambda}\,
\frac{(1\!+\!3\lambda)\nu\!+\!4\pi(2\!-\!3\lambda)\phi^{2}}{(\nu\!+\!8\pi\phi^{2})^{2}}
\dot{\phi}^{ 2}~, \label{egm}
\end{equation}
from where it is seen when it could be possible to have
acceleration, $\ddot{a}>0$. Using equations (\ref{drs}) and
(\ref{rta}), the acceleration (\ref{egm}) can become explicitly a
function solely of $\phi$ given that the equation (\ref{rta}) has
been integrated
\begin{equation}
-\frac{3}{4\pi c^{2}}\frac{\chi^{3}}{\sqrt{|\nu\!+\!8\pi\phi^{2}|}}
\frac{\ddot{a}}{a}=\sqrt{\frac{\epsilon}{\lambda}\!+\!\frac{2\rho_{\ast}}{c^{2}}\chi}
\,\Bigg(\sqrt{\frac{\epsilon}{\lambda}\!+\!\frac{2\rho_{\ast}}{c^{2}}\chi}
\mp\sqrt{\frac{3}{\pi}}\frac{4\pi\epsilon|\phi|}{\sqrt{|\nu\!+\!8\pi\phi^{2}|}}\Bigg)\!+\!
\frac{(1\!+\!3\lambda)\nu\!+\!8\pi\phi^{2}}{\lambda|\nu\!+\!8\pi\phi^{2}|}\,.
\label{eer}
\end{equation}
This expression will be used for the investigation of the
acceleration/deceleration intervals at early times. Accordingly,
for the Brans-Dicke theory, equation (\ref{egs}) takes the form
$\frac{\ddot{a}}{a}=-H^{2}-\frac{k}{a^{2}}-\frac{\omega}{6}\frac{\dot{\phi}^{2}}{\phi^{2}}
$, thus, for $k\geq 0$ and $\omega>0$ there is no acceleration. In
our model, the existence of the parameter $\nu$ gives the freedom
to have acceleration without the need of introducing a
self-coupling of the scalar field or through other mechanisms.
Finally, the Ricci scalar $R$ is found from equations (\ref{egm})
and (\ref{drs}) to have the simple expression
\begin{equation}
R=-\frac{8\pi c^{2}}{\lambda
a^{6}}\,\frac{(1\!+\!3\lambda)\nu\!+\!4\pi(2\!-\!3\lambda)\phi^{2}}
{(\nu\!+\!8\pi\phi^{2})^{2}}~. \label{wef}
\end{equation}

Next, we will discuss the four cases resulting from equation
(\ref{rta}).

\begin{itemize}

\item{
{\textit{\underline{Case I}}}: If $\nu\!+\!8\pi\phi^{2}>0$ and $\epsilon\lambda>0$, then
$\lambda\phi>0$. Equation (\ref{rta}) is integrated to
\begin{equation}
a^{2}=\frac{2c^{2}}{|\lambda|\rho_{\ast}}
\,\frac{1}{\sqrt{\nu\!+\!8\pi\phi^{2}}}\,
\frac{\sigma\Big|4\pi\phi\!+\!\sqrt{2\pi}\sqrt{\nu\!+\!8\pi\phi^{2}}\Big|
^{\!\pm\sqrt{\frac{2}{3|\lambda|}}}}
{\Big[1\!-\!\sigma\big|4\pi\phi\!+\!\sqrt{2\pi}\sqrt{\nu\!+\!8\pi\phi^{2}}\big|
^{\!\pm\sqrt{\frac{2}{3|\lambda|}}}\,\Big]^{2}}~, \label{jei}
\end{equation}
where $\sigma>0$ is integration constant and it should be
$\sigma\big|4\pi\phi\!+\!\sqrt{2\pi}\sqrt{\nu\!+\!8\pi\phi^{2}}\big|
^{\!\pm\sqrt{\frac{2}{3|\lambda|}}}<1$.
We study next the behaviour of this solution for $\phi>0$, which implies
$\epsilon,\lambda>0$.

For $\nu>0$ the upper branch is defined for
$0<\phi<\phi_{1}=\frac{1}{8\pi}
\big(\sigma^{-\sqrt{\frac{3\lambda}{2}}}\!-\!2\pi\nu
\sigma^{\sqrt{\frac{3\lambda}{2}}}\big)$ and note that
$2\pi\nu<\sigma^{-\sqrt{6\lambda}}$. At the minimum value
$\phi=0$, the scale factor $a$ of the solution (\ref{jei}) gets a
constant value with $\rho=0$, thus the Ricci scalar (\ref{wef}) is
also finite. At the maximum value $\phi=\phi_{1}$ the scale factor
becomes infinite and again $\rho\rightarrow 0$. Although the
quadrature (\ref{erv}) of the time cannot be integrated explicitly
in terms of elementary functions, we can find the leading
behaviour close to the minimum value of $\phi$. It is seen that
for $\phi= 0$ the time $t= 0$. Since at the minimum $a$, $\phi$
increases, we must have from (\ref{drs}) that $c>0$, and thus the
function $\phi(t)$ is increasing and it is like a good time
parameter. Therefore, we have a universe which emerges at zero
cosmic time at a finite volume and avoids the cosmological
singularity both in density and curvature.

Equally importantly, there are parameters such that there is a
transient accelerating era in the radiation epoch. For example,
setting $\lambda=2,\nu=0.5,\sigma=0.1,c=3,\rho_{\ast}=1$ in
equation (\ref{eer}) and make a plot of $\frac{\ddot{a}}{a}$ we
see that we get an initial decelerating era, followed by an
accelerated period which finishes, and the universe enters again
into deceleration. This transient acceleration era could be
interpreted as an inflationary period with a graceful exit. It is
also possible, depending on the parameters, to exist a temporary
``breath'' of the scale factor, so $a$ during the expansion, for
some period of time contracts, and then re-expands monotonically.
This phenomenon might deserve further investigation. The
corresponding Brans-Dicke solution arises from equation
(\ref{jei}) for $\nu=0$ and $\lambda>0$. The result is that for
$\lambda<\frac{2}{3}$ the above non-singular branch, where the
universe emerges at finite radius, is lost here, and the solution
becomes singular. For $\lambda>\frac{2}{3}$ ($\omega<0$)
Brans-Dicke possesses a bouncing solution. To give an estimate of
the relative energy densities of the matter and the scalar field
we refer that $\rho/\dot{\phi}^{2}\rightarrow 0$ at the minimum
scale factor and $\rho/\dot{\phi}^{2}\rightarrow \infty$ at
infinity.

For $\nu>0$ the lower branch is defined for $\phi>\text{max}\{0,\phi_{2}\}$,
$\phi_{2}=\frac{1}{8\pi}
\big(\sigma^{\sqrt{\frac{3\lambda}{2}}}\!-\!2\pi\nu
\sigma^{-\sqrt{\frac{3\lambda}{2}}}\big)$, and
it is always decelerating. For $2\pi\nu>\sigma^{\sqrt{6\lambda}}$, the universe starts
at zero $a$ with infinite $\phi$ and infinite $\rho,R$, and finally tends to a constant
scale factor
for $\phi=0$, $\rho=0$. For $2\pi\nu<\sigma^{\sqrt{6\lambda}}$, the universe starts
at zero $a$ with infinite $\phi$ and infinite $\rho,R$, and finally tends to infinite
volume
for $\phi=\phi_{2}$, $\rho=0$. Therefore, these are typical singular solutions.

For $\nu<0$ only the upper branch is valid which is defined for
$\phi_{3}<\phi<\phi_{4}$, where
$\phi_{3}=\sqrt{\frac{|\nu|}{8\pi}}$, $\phi_{4}=\frac{1}{8\pi}
\big(\sigma^{-\sqrt{\frac{3\lambda}{2}}}\!+\!2\pi|\nu|
\sigma^{\sqrt{\frac{3\lambda}{2}}}\big)$ (it should also be
$2\pi|\nu|<\sigma^{-\sqrt{6\lambda}}$). At both $\phi_{3}$,
$\phi_{4}$ the scale factor is infinite and $\rho\rightarrow 0$.
Therefore, this solution describes a bouncing universe where the
universe collapses from infinite volume, it has a bounce with all
$a,\rho$ and $R$ finite, and re-expands to infinity.}

\item{
{\textit{\underline{Case II}}}: If $\nu\!+\!8\pi\phi^{2}<0$ and $\epsilon\lambda<0$, then
$\lambda\phi>0$. Equation (\ref{rta}) is integrated to
\begin{equation}
a^{2}=\frac{c^{2}}{2|\lambda|\rho_{\ast}}\,\frac{1}{\sqrt{|\nu|\!-\!8\pi\phi^{2}}}
\,\Big{\{}1+\tan^{2}\!\Big[\sigma\!\pm\!
\frac{1}{\sqrt{6|\lambda|}}
\arcsin{\Big(\sqrt{\frac{8\pi}{|\nu|}}\,\phi\Big)}\Big]\Big{\}}~,
\label{etk}
\end{equation}
where $\sigma$ is an integration constant and it should be
$0<\sigma\!\pm\!\frac{1}{\sqrt{6|\lambda|}}
\arcsin{\big(\sqrt{\frac{8\pi}{|\nu|}}\,\phi\big)}<\frac{\pi}{2}$.
For $\phi>0$, it is implied $\epsilon<0$, $\lambda>0$.

For the upper branch there are four cases, but they all have the same features.
For $\sigma>0$ and $0<\frac{\pi}{2}-\sigma<\frac{\pi}{2\sqrt{6\lambda}}$ it is
$0<\phi<\phi_{1}$,
$\phi_{1}=\sqrt{\frac{|\nu|}{8\pi}}\sin{[\sqrt{6\lambda}|\frac{\pi}{2}-\sigma|]}
$.
At the minimum value $\phi=0$, the scale factor $a$ of the solution (\ref{etk}) gets a
constant value with $\rho=0$, $R$ finite and $t=0$. At the maximum value $\phi=\phi_{1}$ the scale
factor becomes infinite and also $\rho\rightarrow 0$. Therefore, we have again a non-singular
universe in all the quantities, volume, energy density and curvature. Analogously to the previous
case, here there are parameters such that the evolution starts with acceleration which is followed by
an entrance into deceleration (this happens for example for
$\lambda=0.1,\nu=-0.5,\sigma=0.1,c=3,\rho_{\ast}=1$). The present solution has some
additional interest since it occurs for very small values of $\lambda$ making $\omega>0$. In
addition, it is
valid for negative values of $\nu$ which will be seen in the next section that they provide the
correct phenomenology at late times; therefore, we can have a unified picture for all
times with a unique mechanism of energy transfer between matter and the scalar field.
For $-\frac{\pi}{2\sqrt{6\lambda}}<\sigma<0$ and
$\frac{\pi}{2}-\sigma>\frac{\pi}{2\sqrt{6\lambda}}
$,
it is $\phi_{2}<\phi<\phi_{3}$, where
$\phi_{2}=\sqrt{\frac{|\nu|}{8\pi}}\sin{(|\sigma|\sqrt{6\lambda})}$,
$\phi_{3}=\sqrt{\frac{|\nu|}{8\pi}}$.
For $\sigma>0$ and $\frac{\pi}{2}-\sigma>\frac{\pi}{2\sqrt{6\lambda}}$,
it is $0<\phi<\phi_{3}$.
For $\sigma<0$ and $\frac{\pi}{2}-\sigma<\frac{\pi}{2\sqrt{6\lambda}}$,
it is $\phi_{2}<\phi<\phi_{1}$. In all these cases the behaviour is the same with the
appearance of a finite universe expanding to infinity.

For the lower branch there are also four cases. For $0<\sigma<\frac{\pi}{2}$ and
$\sigma<\frac{\pi}{2\sqrt{6\lambda}}$, it is $0<\phi<\phi_{2}$ and the universe starts
non-singular and ends at a finite scale factor. For
$0<\sigma-\frac{\pi}{2}<\frac{\pi}{2\sqrt{6\lambda}}$ and
$\sigma>\frac{\pi}{2\sqrt{6\lambda}}$,
it is $\phi_{1}<\phi<\phi_{3}$ and the solution represents a bouncing universe.
Finally, for $\frac{\pi}{2\sqrt{6\lambda}}<\sigma<\frac{\pi}{2}$ where it is
$0<\phi<\phi_{3}$,
or for $\frac{\pi}{2}<\sigma<\frac{\pi}{2\sqrt{6\lambda}}$ where $\phi_{1}<\phi<\phi_{2}$,
we have a non-singular universe extending to infinity.}

\item{
{\textit{\underline{Case III}}}: If $\nu\!+\!8\pi\phi^{2}>0$ and $\epsilon\lambda<0$, then
$\lambda\phi<0$. Equation (\ref{rta}) is integrated to
\begin{equation}
a^{2}=\frac{c^{2}}{2|\lambda|\rho_{\ast}}\,\frac{1}{\sqrt{\nu\!+\!8\pi\phi^{2}}}
\Big[1+\tan^{2}\!
\Big(\sigma\!\pm\!\frac{1}{\sqrt{6|\lambda|}}
\ln\Big|4\pi\phi\!+\!\sqrt{2\pi}\sqrt{\nu\!+\!8\pi\phi^{2}}\Big|\Big)\Big]\,,
\label{wkt}
\end{equation}
where $\sigma$ is integration constant and it should be
$0<\sigma\!\pm\!\frac{1}{\sqrt{6|\lambda|}}
\ln\Big|4\pi\phi\!+\!\sqrt{2\pi}\sqrt{\nu\!+\!8\pi\phi^{2}}\Big|<\frac{\pi}{2}$.
For $\phi>0$, it is implied $\epsilon>0$, $\lambda<0$.

For $\nu>0$ the upper branch possesses two cases. The first case
is characterized by the conditions $\phi_{1}<0$ and $\phi_{2}>0$,
where $\phi_{1}=\frac{1}{8\pi}(e^{-\sigma\sqrt{6|\lambda|} }
-2\pi\nu e^{\sigma\sqrt{6|\lambda|}})$,
$\phi_{2}=\frac{1}{8\pi}\big[e^{\sqrt{6|\lambda|}(\frac{\pi} {2}
-\sigma)}-2\pi\nu
e^{-\sqrt{6|\lambda|}(\frac{\pi}{2}-\sigma)}\big]$, which means
equivalently
$e^{-2\sigma\sqrt{6|\lambda|}}<2\pi\nu<e^{2\sqrt{6|\lambda|}(\frac{\pi}{2}-\sigma)}$.
Then it is $0<\phi<\phi_{2}$. At the minimum value $\phi=0$ the
universe is non-singular, with finite $a$, $\rho=0$ and $R$
finite. At the maximum value $\phi=\phi_{2}$ the scale factor
extends to infinity with $\rho\rightarrow 0$. Moreover, there are
parameters such that the universe starts decelerating, there is a
transient accelerating era and an exit to deceleration (for
example this happens for
$\lambda=-0.5,\nu=0.5,\sigma=0.1,c=3,\rho_{\ast}=1$). The second
case refers to $0<\phi_{1}<\phi_{2}$ and it is
$\phi_{1}<\phi<\phi_{2}$. Again at the minimum $\phi=\phi_{1}$ all
$a,\rho$ and $R $ are finite, while at the maximum
$\phi=\phi_{2}$, the volume becomes infinite.

For $\nu>0$ the lower branch has also two cases. For the first it is $\phi_{4}<0$,
$\phi_{3}>0$,
where $\phi_{3}=\frac{1}{8\pi}(e^{\sigma\sqrt{6|\lambda|}}-2\pi\nu
e^{-\sigma\sqrt{6|\lambda|}})$,
$\phi_{4}=\frac{1}{8\pi}\big[e^{\sqrt{6|\lambda|}(\sigma-\frac{\pi}{2})}
-2\pi\nu e^{-\sqrt{6|\lambda|}(\sigma-\frac{\pi}{2})}\big]$ and thus $0<\phi<\phi_{3}$. At
the
minimum
$\phi=0$ the universe starts non-singular and ends for $\phi=\phi_{3}$ also to a finite
universe.
For the second case it is $0<\phi_{4}<\phi_{3}$ and thus $\phi_{4}<\phi<\phi_{3}$.
Again we have a non-singular universe extending to infinity.

For $\nu<0$ there is a multiplicity of cases for each branch and
we get analogous behaviours as those mentioned above.}

\item{
{\textit{\underline{Case IV}}}: If $\nu\!+\!8\pi\phi^{2}<0$ and $\epsilon\lambda>0$, then
$\lambda\phi<0$. Equation (\ref{rta}) is integrated to
\begin{equation}
a^{2}=\frac{c^{2}}{2|\lambda|\rho_{\ast}}\,\frac{1}{\sqrt{|\nu|\!-\!8\pi\phi^{2}}}\,
\sinh^{-2}\!\Big[\sigma\mp\frac{1}{\sqrt{6|\lambda|}}
\arcsin{\Big(\sqrt{\frac{8\pi}{|\nu|}}\,\phi\Big)}\Big]\,,
\label{bhj}
\end{equation}
where $\sigma$ is integration constant and it should be
$\sigma\mp\frac{1}{\sqrt{6|\lambda|}}
\arcsin{(\sqrt{\frac{8\pi}{|\nu|}}\,\phi)}>0$. For $\phi>0$, it is implied $\epsilon<0$,
$\lambda<0$.

For the upper branch it should be $\sigma>0$ and $0<\phi<\phi_{1}$, where
$\phi_{1}=\sqrt{\frac{|\nu|}{8\pi}}\sin\big(|\sigma|\sqrt{6|\lambda|}\big)$.
At the minimum $\phi=0$ we get a non-singular universe with $a$ finite, $\rho=0$ and $R$
finite.
At the maximum $\phi=\phi_{1}$ the scale factor extends to infinity and $\rho\rightarrow
0$.
We can easily get decelerating solutions in this branch.

For the lower branch there are two cases. The first one has
$\sigma>0$ and $0<\phi<\phi_{2}$, where
$\phi_{2}=\sqrt{\frac{|\nu|}{8\pi}}$. For $\phi=0$ the universe is
non-singular with finite all $a,\rho$ and $R$, while for
$\phi=\phi_{2}$ the scale factor goes to infinity with
$\rho\rightarrow 0$. We can check numerically that there is an
intermediate contracting phase. Initially we can have a
decelerating phase, followed by an eternal acceleration. The
second case has $\sigma<0$ and $\phi_{1}<\phi<\phi_{2}$. At both
$\phi_{1},\phi_{2}$ the scale factor $a\rightarrow \infty$ and the
solution represents a bouncing universe.}

\end{itemize}

To resume with the most interesting solutions, for $\nu>0$ (with either $\lambda>0$ - Case
I or
$\lambda<0$ - Case III) we can have an avoidance of the initial singularity (in all, scale
factor,
proper time, energy density and curvature) and at the same time a transient accelerating
period
within deceleration in the radiation regime. For $\nu<0$ (such that
$\nu\!+\!8\pi\phi^{2}<0$)
and $\lambda>0$ - Case II we can have again a non-singular universe in all the quantities,
where the
evolution starts with acceleration which is followed by an entrance into deceleration.

In all these cases, due to the interaction term on the right-hand
side of equation (\ref{euf}), we have an entropy production which
could help to confront the cosmological entropy problem. Namely, we make
use of the standard thermodynamic relation $dU+pdV=TdS$, where
$U=\rho V$ is the energy contained in a comoving volume $V\propto
a^{3}$ with corresponding entropy $S$ and temperature $T$. Then,
equation (\ref{euf}) can be written as
\begin{equation}
\frac{T}{V}\dot{S}=\frac{\nu}{\phi(\nu\!+\!8\pi\phi^{2})}\rho\dot{\phi}\,.
\label{hyw}
\end{equation}
Whenever the right-hand side of equation (\ref{hyw}) is positive,
as it happens with the branches mentioned above, the universe
evolution leaves adiabaticity and leads to entropy production.
This entropy is shared initially between all relativistic species (photons, baryons, etc.).
But as the universe cools down, the massive particles freeze out and the entropy is only
shared to the photons. These photons propagate in the universe and are observed today with their
high value of entropy per baryon, while the corresponding temperature $T$ scales as $1/a$.
Of course there is the constraint that the mechanism which produces entropy and matter, should not
create too much matter, in order to comply with observations.

\section{Late-times cosmological evolution}

In this section we will investigate the late-times cosmology of
the generalized Brans-Dicke gravity. The Friedmann equations
(\ref{eji}) and (\ref{ket}) can be  written in a more familiar
form
\begin{eqnarray}
&&H^{2}+\frac{k}{a^{2}}=\frac{8\pi}{3\phi}\left(\rho+\rho_{DE}\right)
\label{FR1}\\
&&2\dot{H}+3H^{2}+\frac{k}{a^{2}}=-\frac{8\pi}{\phi}\left(p+p_{DE}\right)\,,
\label{FR2}
\end{eqnarray}
where we have defined the effective dark energy and effective dark
pressure as
\begin{eqnarray}
&&\rho_{DE}\equiv -\frac{3\phi^2}{\nu\!+\!8\pi\phi^{2}} H\dot
{\phi}
+\frac{\phi}{2\lambda}\,\frac{\nu\!+\!4\pi(2\!-\!3\lambda)\phi^{2}}
{(\nu\!+\!8\pi\phi^{2})^{2}}\dot{\phi}^{2}
\label{rhoDE}\\
&&p_{DE}\equiv
\frac{\phi}{2\lambda}\,\frac{(1\!+\!2\lambda)\nu\!+\!4\pi(2\!-\!3\lambda)\phi^{2}}
{(\nu\!+\!8\pi\phi^{2})^{2}}
\dot{\phi}^{2}+\frac{\phi^{2}}{\nu\!+\!8\pi\phi^{2}}(2H\dot{\phi}+\ddot{\phi})~.
\label{pDE}
\end{eqnarray}
Observe that we recover the standard Brans-Dicke cosmology if
$\nu=0$. If $\nu\neq0$ the Friedmann equations (\ref{FR1}) and
(\ref{FR2}) incorporate in a non-trivial way the time evolution of
the scalar-field which, as we will discuss in the following,
brings new features in the late-times cosmological evolution
compared to the standard Brans-Dicke theory.

To study the late-times cosmology we define the equation-of-state
parameter for the effective dark energy sector
\begin{eqnarray}
w_{DE}\equiv \frac{p_{DE}}{\rho_{DE}}~, \label{wDE}
\end{eqnarray}
and we introduce the deceleration parameter through
\begin{eqnarray}
 q=-1-\frac{\dot{H}}{H^{2}}~.
\label{qq}
\end{eqnarray}
Then, according to (\ref{FR1}), we can define the density
parameters as
\begin{equation}
\Omega_{m}=\frac{8\pi\rho}{3\phi
H^2}~,\,\,\,\,\,\,\,\,\,\Omega_{DE} =\frac{8\pi\rho_{DE}}{3\phi
H^2}~. \label{qed}
\end{equation}
Thus, the deceleration parameter (\ref{qq}) can be written for
$k=0$ in terms of the density parameters  as
\begin{equation}
 q=\frac{1}{2}+\frac{3}{2}\big(w_{m}\Omega_{m}+w_{DE}\Omega_{DE}\big)~,
\end{equation}
with $w_{m}\equiv p/\rho$ the matter equation-of-state parameter.

If we combine equations (\ref{elp}) and (\ref{idj}), we get the
conservation equation of $\rho_{DE}$
\begin{equation}
\dot{\rho}_{DE}+3H(\rho_{DE}+p_{DE})=\Big(\frac{1}{\phi}\rho_{DE}+
\frac{8\pi\phi}{\nu\!+\!8\pi\phi^{2}}\rho\Big)\dot{\phi}~.
\label{egr}
\end{equation}
Additionally, we can rewrite the conservation equations
(\ref{euf}) and (\ref{egr}) in the form
\begin{eqnarray}
&&\Big(\frac{1}{\phi}\,\rho\Big)^{\cdot}+\frac{3H}{\phi}(\rho+p)
=-\frac{8\pi}{\nu\!+\!8\pi\phi^{2}}\rho\dot{\phi}=-Q
\label{drg}\\
&&\Big(\frac{1}{\phi}\,\rho_{DE}\Big)^{\cdot}+\frac{3H}{\phi}(\rho_{DE}+p_{DE})
=\frac{8\pi}{\nu\!+\!8\pi\phi^{2}}\rho\dot{\phi}=Q\label{asd}~.
\end{eqnarray}
The above relations (\ref{drg}) and (\ref{asd}) resemble the
standard relations
\bea && \dot{\rho}+3H(\rho+p)=-Q \nonumber \\
&& \dot{\rho}_{DE}+3H(\rho_{DE}+p_{DE})=Q~,
\eea
which give rise to an interaction between the effective dark energy and the
dark matter sector, with the quantity $Q$ giving the
strength and the form of this interaction. However, there are two
main differences. The first one is that the scalar field $\phi$,
which plays the role of a varying gravitational constant, enters
the variation too and hence, the quantity that is strictly
conserved is $(\rho+\rho_{DE})/\phi$. The second one is that the
above interaction, and the specific expression for $Q$, is
determined by the theory itself through the consistency of the extension of
the standard Brans-Dicke theory.

In the existing literature, the interaction term $Q$ is only
phenomenological and it is not resulted from a consistent field
theoretic model. In \cite{Bean:2008ac}, for example, this coupling
parameterizes the interaction between dark matter to a
quintessence field and it is constrained from observations
compared to the couplings of the fields to gravity. In
\cite{Micheletti:2009pk} an interactive field theory is discussed
in which the interaction $Q$ describes the coupling of a fermionic
field for dark matter to a bosonic field for dark energy. In the
model discussed here, this interactive term results from the theory, and as can be seen
from equation (\ref{drg}), for $\nu=0$ it recovers the
standard conservation equation (\ref{eue}) of non-interacting
Brans-Dicke cosmology. The dependence of the interactive term $Q$
on the parameter $\nu$ can be understood from the fact that from
equation (\ref{idj}) or (\ref{euf}) the matter energy-momentum
tensor is sourced by the scalar field, where the interaction
is controlled by the parameter $\nu$. Similarly, in
equation (\ref{lrs}) or (\ref{iet}) it is seen that the matter
energy-momentum tensor is the source of the scalar field and the
strength of the interaction is now controlled by $\lambda$.

The main motivation for the study of interacting models is to solve the coincidence
problem with a suitable coupling between dark energy and dark matter and also to drive
the transition from an early matter dominated era to a phase of accelerated expansion
\cite{Amendola:1999qq}. The important issue which is known is that there is the
possibility for a scaling attractor to be accelerating only if the standard conservation of matter is
violated. Another motivation is the prediction of the variation of $w_{DE}$
during the evolution of the universe. The variation of $w_{DE}$
was studied in \cite{Wang:2005jx} using an appropriate coupling
between dark energy and dark matter and it was shown in a
holographic model that a transition from $w_{DE}>-1$ to
$w_{DE}<-1$ occurs, claiming that this property could serve as an
observable feature of the interaction between dark energy and dark
matter, in addition to its influence on the small $l$ Cosmic Microwave Background spectrum
argued in \cite{Zimdahl:2005bk}.

Concerning equations (\ref{drg}), (\ref{asd}) there are strong constraints
about the possible non-gravitational interactions of baryons. In most of the works
on interacting dark energy models, the baryonic matter is separately conserved
and only dark matter is allowed to interact non-gravitationally with dark energy.
This would mean that the parameter $\nu$ corresponding to the conservation equation of
the baryonic matter is restricted to very small values, so that the weak equivalence
is respected by this component.
In this work we do not consider such a distinction between the two components (dark/baryonic)
since we do not attempt a fitting against real data and our analysis is merely
indicative. Our main results, however, are not expected to change significantly under a
more detailed analysis since the baryonic matter forms a small percentage of the total one.

To study the two Friedmann equations (\ref{FR1}) and (\ref{FR2})
we will ignore at first approximation the radiation
component and consider that the universe is composed of
non-relativistic matter $\rho$ with negligible pressure
$p\ll\rho$, thus we set $w_{m}=0$.  Even in this case, the
complexity of equations (\ref{rhoDE}) and (\ref{pDE}) does not
allow us to have analytical solutions, and hence in the following
we will use numerical methods.  In order to acquire a consistent
cosmology, in agreement with observations, we restrict to the flat
case, namely imposing $k=0$, and we set the present values of the
density parameters to $\Omega_{m0}\approx0.3$ and
$\Omega_{DE0}\approx0.7$ \cite{Ade:2013sjv}. In addition, we set
the standard value $H_{0}=0.71\times 10^{-10}\,\text{yr}^{-1}$ and
we can assume for convenience that the present scale factor is
$a_0=1$ and the units are chosen so that $\phi_0=1$. These values
allow us to determine from equation (\ref{rhoDE}) and the second
of equations (\ref{qed}) the present value $\dot{\phi}_{0}$ as a
function of the parameters $\lambda,\nu$.

Note that since the expression of $\rho_{DE}$ in equation
(\ref{rhoDE}) is quadratic in $\dot{\phi}$, the above procedure
will give two values for $\dot{\phi}_{0}$, and each one will give rise
to a different cosmological evolution. Additionally, there can be
regions of the model parameters $\lambda$, $\nu$ for which
$\dot{\phi}_{0}$ becomes complex, and thus the corresponding
cosmologies are not realistic. Next, combining equation
(\ref{keg}) with the first of relations (\ref{qed}), and applying
them on present time, we obtain (of course $\phi$ is positive)
\begin{equation}
\rho_{\ast}=\frac{3a_0^3\Omega_{m0}H_{0}^{2}}{8\pi}\sqrt{|\nu\!+\!8\pi\phi_0^2|}~.
\label{rstarH0}
\end{equation}
This equation provides  the integration constant $\rho_{\ast}$
defining the evolution of the energy density $\rho$. Having chosen
the initial conditions $a_{0}, \phi_{0}, \dot{\phi}_{0}$ as above,
we are going to solve numerically the system of equations
(\ref{eji}) and (\ref{iet}) for $a(t),\phi(t)$ with the function
$\rho$ given by equation (\ref{keg}). After these solutions have
been obtained, we can derive the phenomenological quantities
$\rho_{DE},  \Omega_{m}, \Omega_{DE}, w_{DE}, q$  as functions
of time or of the redshift $z$. Having satisfied the present cosmological data,
we can explore the cosmological behaviour for various
values of the model parameters $\nu$ and $\lambda$, requiring a
realistic cosmology, namely a present dark energy
equation-of-state parameter $w_{DE0}$ around $-1$ and a matter
domination at earlier times.
\begin{figure}[ht]
\begin{center}
\includegraphics[width=0.7\textwidth]{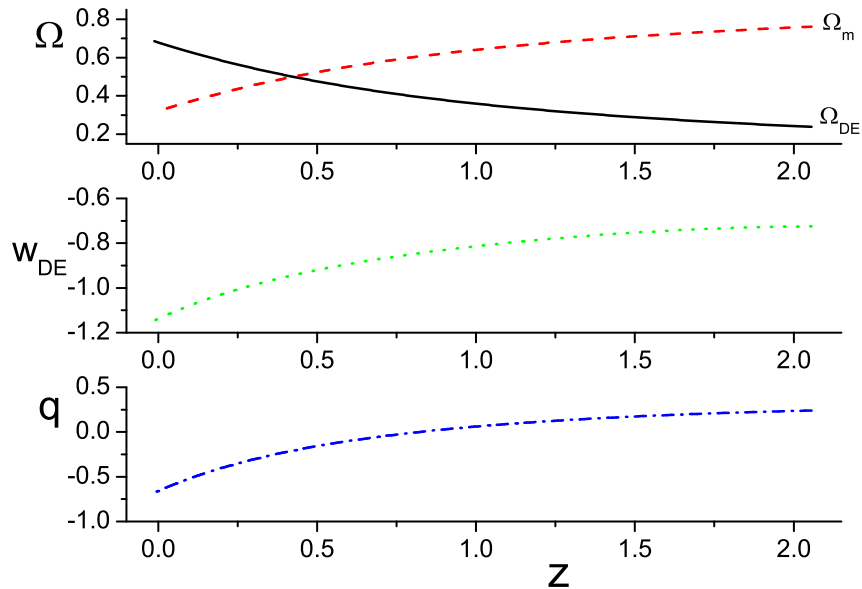}
\caption{{\it{The late-times cosmological evolution for a spatially flat
universe, for the parameter choice $\nu=-100$ and $\lambda=10$ in
units where $\phi_0=1$, having imposed $\Omega_{m0}\approx0.3$,
$\Omega_{DE0}\approx0.7$ at present, and having set the present
scale factor $a_0=1$. As independent variable we use the redshift
$z=-1+a_0/a$. In the upper graph we depict the evolution of the
matter and dark energy density parameters. In the middle graph we
present the evolution of the dark energy equation-of-state
parameter $w_{DE}$. In the lower graph we depict the evolution of
the deceleration parameter $q$.}}} \label{basicplot}
\end{center}
\end{figure}

In Fig. \ref{basicplot} we present the cosmological evolution for
a spatially flat universe for the parameter choice $\nu=-100$ and
$\lambda=10$. As independent variable we use the redshift
$z=-1+a_0/a$. In the upper graph we depict the evolution of the
matter and dark energy density parameters, which their behaviour
show an agreement with observations. In the middle graph we
present the evolution of the dark energy equation-of-state
parameter $w_{DE}$. Finally, in the lower graph we depict the
evolution of the deceleration parameter $q$, where  the passage
from deceleration to acceleration at late times can be seen.

Note that in the above figures we have checked that $\phi(t)$
exhibits an increasing behaviour up to very large redshifts which
is the result of the fact that we have chosen $\nu$ and $\lambda$
in order for $\dot{\phi}_0$ to be positive. This means that the
gravitational constant $G$ is decreasing with time as it is expected
to be more reasonable in Brans-Dicke theory. However, $\phi$
does not cross the zero value which would mean a discontinuity
from $+\infty$ to $-\infty$ for $G$. Additionally, we have checked
that $\frac{\dot{\phi}}{\phi H}\big|_{0} \lesssim 10^{-2}$ which
is necessary in order to be consistent with the bounds of the
variation of the Newton's constant. The redshift at which acceleration starts,
has been reasonably determined observationally to
$z=0.74\pm 0.05$ \cite{Farooq:2013hq}, and might set non-redundant constraints
on the parameter $\nu$. The product of the age of the universe $t_{0}$ with the current
Hubble constant $H_{0}$ is $0.81\lesssim t_{0}H_{0}\lesssim 1.09$ \cite{Ade:2013sjv},
\cite{Chaboyer:1998fu}-\cite{Ade}
and might also set constraints on the parameters. We will not consider these two bounds here.

As it is well known the standard Brans-Dicke theory cannot give a
consistent cosmology, and in particular it cannot lead to
acceleration for realistic values of $\omega$. To remedy this, a
potential for the scalar field was introduced which is beyond the
original construction of the Brans-Dicke theory because this potential term drives the
acceleration and not the Brans-Dicke kinetic terms.  Also other
alternatives were assumed, as discussed in the Introduction. In
the extended Brans-Dicke model studied here, the late cosmological evolution is in agreement with
observations and in particular the new parameter $\nu$, which
arises from the generalization of the standard Brans-Dicke
theory, can drive alone the acceleration without the need of a
potential.

%Additionally, we mention that in the original Brans-Dicke theory a
%slight variation of $\phi$ seems problematic to be realized due to
%the absence of the extra parameter $\nu$.

Generalizing Fig. \ref{basicplot}, there is a range of the parameters
$\lambda>0$ and $\nu$ sufficiently negative such that $\nu\!+\!8\pi\phi^{2}<0$, which provide
a late-times cosmological evolution in agreement with observations. Such values of $\lambda,\nu$
can match with Case II of the early-times cosmic evolution discussed in (\ref{etk}).
Thus for this parameter region, there is a unified description of the universe cosmic history
to account for inflation, matter domination and late-times acceleration. In addition, from
the conservation equation (\ref{euf}), since $\dot{\phi}>0$, it is seen that $\dot{\rho}$,
beyond the standard dilution, gets a positive contribution, which
means that there is an energy transfer from the scalar field to
the dark matter.

For completeness, we mention that one may also obtain cosmological
evolution in agreement with observations, namely similar to Fig.
\ref{basicplot}, in the case where $\lambda<0$, $\nu$ negative but
not sufficiently, i.e. with $|\nu|\sim 8\pi\phi^{2}$ and $\nu\!+\!8\pi\phi^{2}>0$.
Now, there is an energy transfer from
the dark matter sector to the dark energy component. On the other
hand, for $\nu<0$ with $|\nu|<<8\pi\phi^{2}$ (which is a slight deviation of Brans-Dicke),
or for $\nu\geq0$ (which includes the usual Brans-Dicke with $\nu=0$)
a cosmological evolution in agreement with observations is impossible
(one might actually get acceleration but with increasing $\Omega_m$).

\begin{figure}[ht]
\begin{center}
\includegraphics[width=0.7\textwidth]{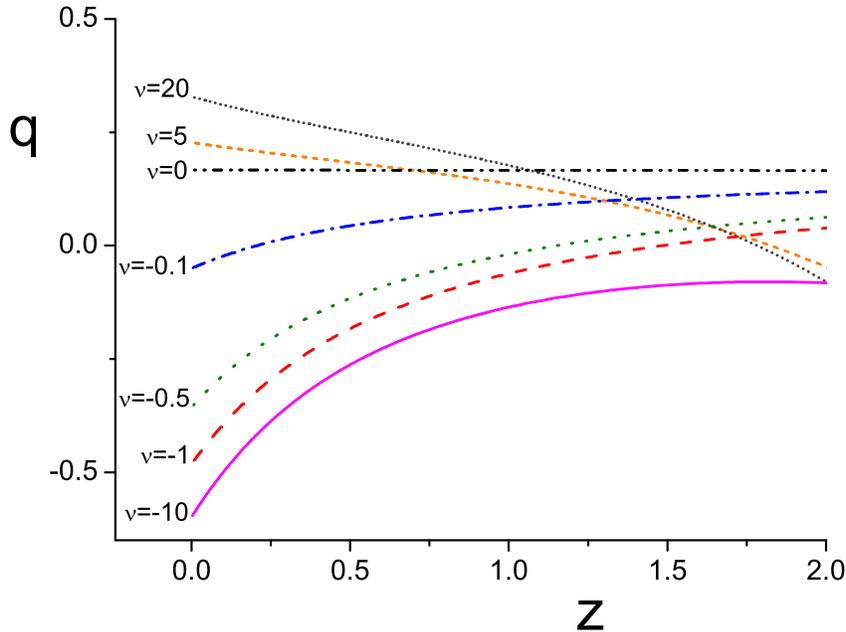}
\caption{{\it{The evolution of the deceleration parameter $q$
versus the redshift $z$ for a spatially flat universe, for
$\lambda=-10$ and for various values of $\nu$.}}} \label{plotq}
\end{center}
\end{figure}

One comment should be added at this point. As we have seen, for $\nu<0$ consistent cosmologies can
arise. However, for such $\nu$'s, equations (\ref{eji}), (\ref{ket}), (\ref{euf}) present a pole
for the value $\phi^{2}=-\nu/(8\pi)$. It is reasonable to wonder if this divergence propagates
any divergence in some quantity at finite time, such as the scale factor, Hubble parameter e.tc.
The answer is negative. For most of the parameter values there is no divergence in any quantity.
This has been checked from the far future up to a high redshift $z=3600$ where the validity of the
dust solution terminates and the radiation solution starts.
The reason for this smooth behaviour is that the scalar field is at all times away from the pole,
so the pole lies outside the regime of applicability of the solution.
We note however that, for some
parameter regions, such as $\nu+8\pi \phi^2<0$ with $\phi$ increasing, the pole value
$\phi^{2}=-\nu/(8\pi)$ might be reached in the future (unless if one
chooses the initial conditions for $\phi$ so that it takes an asymptotically
constant value away from the pole). In this case, the scale factor becomes infinite at a finite
time, which is just the realization of a Big Rip (e.g. \cite{Caldwell:2003vq}). This is expected to
happen in all models where super-acceleration is achieved and the dark energy
equation-of-state parameter lies in the phantom regime.

In Fig. \ref{plotq} we present the evolution of the deceleration parameter $q$ for
various values of $\nu$ and for fixed $\lambda$. We consider
$\lambda<0$ and the two regions: $\nu<0$ with
$\nu\!+\!8\pi\phi^{2}>0$, and $\nu>0$. As mentioned above, for intermediate negative values
of $\nu$ we obtain late-times acceleration. However, for positive $\nu$, as
well as for $|\nu|$ very small, late-times
acceleration cannot be obtained. This behaviour verifies the basic
advantage of the present model that a non-zero parameter $\nu$ can
lead to new and qualitatively different results that can make
Brans-Dicke theory in agreement with observations without the need
of introducing an arbitrary potential or other ingredients.

In summary, in the extended Brans-Dicke theory discussed here, one obtains in cosmology an
interaction between the dark matter and the effective dark energy sectors,
which leads to a late-times cosmological evolution in agreement
with observations.

\section{Conclusions} \label{Conclusions}

We consider a completion of Brans-Dicke theory presented in \cite{Kofinas:2015nwa}, derived
assuming that the scalar field follows a simple massless wave equation sourced by the trace
of the matter energy-momentum tensor, while at the same time the scalar field energy-momentum
tensor contains terms with two derivatives each. The conservation equation of matter
is modified through a specific interaction with the scalar field determined
by the theory and parametrized by a new dimensionfull parameter $\nu$. For $\nu=0$ the
standard Brans-Dicke theory is recovered. In the present work we investigated the
cosmological implications of this theory.

Considering a spatially homogeneous and isotropic geometry we
derived the Friedmann equations of the theory in which non-trivial
field kinetic terms appear similar to the Brans-Dicke gravity, but
with not trivial $\phi$ dependence. Additionally, the standard
conservation equation of a perfect fluid gets extra contribution
from its interaction with the scalar field. This non-conservation
equation can be integrated and the conventional dilution of the
matter energy density gets corrected by a $\phi$-dependent factor.
Significant modifications of the energy density evolution occur
when the scalar field gets values comparable or smaller than the
new mass scale $\nu$.

In the radiation era the system has been reduced to a first order differential equation
for the scale factor as a function of the scalar field. This equation has been integrated and the time
dependence comes through a quadrature. Among the solutions found, there are branches of solutions for
a range of parameters which avoid the initial singularity (in all, scale factor, proper
time, energy density and curvature) and at the same time they possess in the radiation regime either a
transient accelerating period within deceleration, or the evolution emanates with acceleration which
is followed by an entrance into deceleration. Additionally, due to the interaction term in
the non-conservation equation of matter, these branches lead to an entropy production, which
remains to be investigated if can answer the cosmological entropy problem.

In the dust era, we have integrated the system numerically and found late-times
acceleration in agreement with the correct behavior of the density parameters and the dark energy
equation of state. Moreover, the scalar field $\phi$ appears only a slight variation over a large
redshift interval, which leads to agreement with the strict bounds on the variation of the
gravitational constant $G$. There are parameters of the model which provide a unified description of the
universe history, namely account for inflation, matter domination and late-times
acceleration under the same mechanism of energy transfer between matter and the scalar field.
The problem of the nature of the attractors of the theory which is related to
the coincidence problem needs a separate investigation, as well as the complete
confrontation with real data.

\[ \]
{{\bf Acknowledgements}}
The authors would like to thank Elcio Abdalla, Gilles Esposito-Farese, Diego Pavon and Bin Wang
for reading the manuscript and for their very helpful comments and remarks.

%\appendix

%\section{Geometric Components} \label{geometric components}

%%%%%%%%%%%%%%%%%%%%%%%%%%%% BIBLIOGRAPHY %%%%%%%%%%%%%%%%%%%%%%%%%%%%%%%%%%%%%%

\end{document}